\documentstyle[12pt]{article}

\def \pcm3{ ${\rm cm}^{-3}$}
\def \up2{$^{-2}$}
\def \deg{$^{\rm o}$}
\def \ion{$^{+}$}
\def \kms{${\rm km}~{\rm s^{-1}}$}
\begin{document}
\baselineskip 14pt

\begin{center}
{\large \bf The Galactic Environments of Nearby Cool Stars} \\
{\bf Priscilla C. Frisch} \\
{\it University of Chicago; Dept. Astronomy \& Astrophysics} \\
\end{center}

\begin{center}
{\bf Abstract}
\end{center}

The definition of nearby star systems is incomplete without
an understanding of the dynamical interaction between the stars
and ambient interstellar material.  The Sun itself has been 
immersed in the Local Bubble interior void for millions of years, and
entered the outflow of interstellar material from the Scorpius-Centaurus
Association within the past $\sim$10$^5$ years. Heliosphere
dimensions have been relatively large during this period.
A subset of nearby stars have similar recent histories, and astrosphere
properties are predictable providing ambient interstellar matter 
and stellar activity cycles are understood.  The properties of astrospheres
can be used to probe the interstellar medium, and in turn outer planets 
are more frequently immersed in raw interstellar material than inner planets.

\begin{center}
{\bf Astrospheres of Nearby Stars}
\end{center}
Interstellar matter (ISM) governs the interplanetary environment of nearby cool
star systems since neutral interstellar gas penetrates stellar wind bubbles
(astrospheres).
In the case of the heliosphere (the solar wind bubble around the Sun),
98\% (by number) of the diffuse gas in the heliosphere is interstellar gas.
By analogy with the heliosphere, stellar astrospheres can be modeled by equating the ram pressure of the
stellar wind (which depends on activity cycles) and the ram pressures 
of the surrounding interstellar cloud.
Heliosphere models explain many observable particle populations and
phenomena seen by spacecraft, such as
the distribution and ionization of interstellar neutrals, the
pickup ion and anomalous cosmic ray daughter products,
and the distribution of interstellar dust (see Landgraf paper, this volume).
Outer planets are more likely to be immersed in raw interstellar material
than inner planets.

The interstellar pressure on an astrosphere is set by
charged ISM components 
which are excluded from the astrosphere by the Lorentz 
force -- interstellar ions (including those formed by charge exchange
in the astropause regions),
low energy cosmic rays, and the smallest  interstellar dust grains.
Astrosphere dimensions for several nearby G-stars have been 
estimated and are shown in the attached table (from Frisch 1993).  
For example, our nearest star the Sun has a
heliosphere radius of $\sim$120 AU, and a weak  bow shock may be located at
$\sim$200 AU.  Heliosphere properties vary with both ambient interstellar properties
and with the solar activity
cycle, of which the well-known Forbush decrease in cosmic-ray intensity
is an example.

\begin{figure}[t!]
\vspace*{1in}
\includegraphics{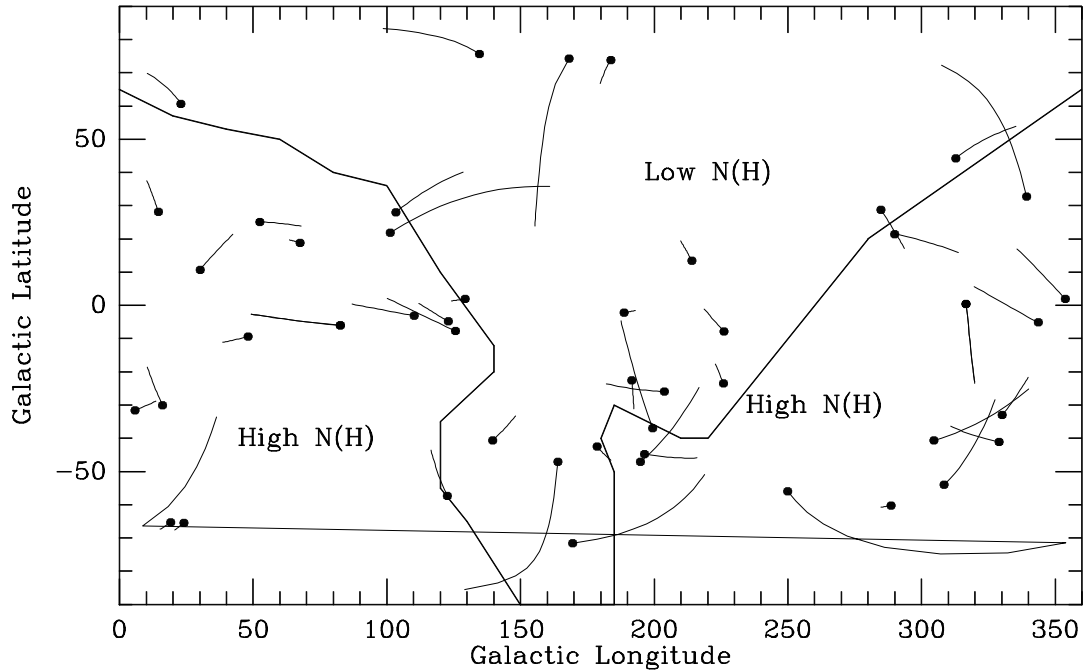}
\vspace*{1.8in}
\caption{Space trajectories of bright stars within 10 parsecs, over
timescales of 0.8 10$^5$ years.  Dots are current star positions.
Regions of high and low density interstellar matter are denoted,
based on results from Genova et al. (1990, ApJ, v355, p150).
Note that stars located in the galactic center hemisphere, and at low galactic
latitudes, are more likely to be immersed in clouds
yielding accretion of ISM onto planetary atmospheres.}
\end{figure}

\begin{center} {\bf Local Interstellar Matter} \end{center}

We have a basic understanding of the properties of ISM within 25 pc of the Sun. 
The interstellar cloud surrounding the solar system is warm, partially
ionized and low density (T$\sim$7000 K, n(H\deg)$\sim$0.22 \pcm3, n(H\ion)$\sim$0.1 \pcm3 (e.g. Frisch et al. 1999).  
The relative cloud-Sun velocity is $\sim$26 \kms.
The thumbprint of 100--200 \kms interstellar shocks is apparent 
from the enhanced abundances of refractory elements observed in nearby ISM,
resulting from interstellar grain destruction (Frisch et al. 1999).  
The distribution of nearby (d$<$30 pc) interstellar material is highly asymmetric,
with the bulk of material located in the galactic-center hemisphere
and low galactic latitudes (see Fig. 1).
Local ISM is structured, indicating inhomogeneous densities.
If a density clump of n(H\deg)=10 \pcm3 were embedded in
the cloud surrounding the solar system and encountered by the Sun, the
heliosphere radius would shrink to about 15 AU (from the current $\sim$120 AU) and
the heliopause would become unstable.  The mass density of interstellar neutrals
would increase from $\sim$1 x 10$^{-25}$ to $\sim$5 x 10$^{-24}$ g \pcm3 at the 1 AU location of the Earth after
such an encounter, dramatically altering the Earth's interplanetary
environment (see Zank and Frisch 1999).
Outer planets are therefore more likely to be exposed to raw interstellar
material then inner planets.

The outflow of interstellar material from the nearby Scorpius-Centaurus
Association governs the galactic environment of the Sun and other
nearby stars.  This author believes that the Sun is immersed
in the leading edge of a superbubble shell associated with the
latest epoch of star formation in the Scorpius-Centaurus Association
(Frisch 1998b).
Velocities of nearby ISM clouds cluster about a 
vector motion consistent with a gas flow from the Scorpius-Centaurus 
Association.  
A bulk flow velocity in the LSR of --20 \kms from the direction 
l=315\deg, b=--3\deg provides a better match to radial velocities of interstellar
clouds observed towards nearby stars than does the
LSR "standard" velocity frame.
(This value for the LSR upwind direction depends on the value used for
the solar apex motion; a recent apex velocity based on Hipparchos
data yields an interstellar flow velocity of
V=--15 \kms, arriving from l=344\deg, b=--2\deg, Frisch 1999).   

\begin{center} {\bf Paleoastrospheres} \end{center}

The historical astrosphere of a star can be predicted by comparing the stellar
space trajectory with the distribution and motions of interstellar clouds.
Space motions of stars can be extrapolated back in time for several million
years.  The dynamics of interstellar clouds are governed by
star formation activity (for diffuse clouds) and spiral arm patterns
(for molecular clouds).
The ISM is highly structured, with tenuous hot plasmas and warm diffuse low
density material both yielding large astrospheres (although the interplanetary
environments differ for these two cases).
Astrosphere properties can be predicted based on properties of interstellar clouds
surrounding each star, and relative to space motions of the star and ISM.  
The following nearby stars are predicted to have astrospheres unchanged over the
past several million years, with astropause radii $\sim$65-75 AU, based on the space
trajectories of each star (from Frisch 1993): 

\begin{table}[t]
\caption{Nearby Star Astrospheres}
\begin{center}
\begin{tabular}{rllcccc} 
\hline
HD &  Name &  Spec &  LSR Total Velocity$^{\rm (a)}$ & Long.& Lat. & dist \\
 &            &           &                     (\kms) &  (deg) & (deg) & (pc) \\
\hline
13421 &  64 Cet &  G0 IV &  39 &  155 &  -49 &  30\\
14412 &   &  G5 V &  34 &  214 &  -70 &  12\\
48938 &   &  G2 V &  35 &  237 &  -13 &  17\\
50692 &  37 Gem &  G0 V &  25 &  191 &  13 &  19\\
84737 &  15 LMi &  G0 V &  26 &  173 &  50 &  13\\
147513 &   &  G5 V &  23 &  342 &  7 &  15\\
181655 &   &  G8 V &  27 &  70 &  11 &  24\\
\end{tabular}
\end{center}
($^{\rm a)}$ The absolute space velocity of the stars in the LSR
(i.e.(V$_{\rm x}^{\rm 2}$ + (V$_{\rm y}^{\rm 2}$ + (V$_{\rm z}^{\rm 2}$)$^{\rm 1/2}$), with
using a solar apex motion for conversion to the LSR of 16.5 km/s towards
l=53\deg, b=25\deg.
\end{table}

\newpage
\begin{center} {\bf Conclusions} \end{center}

The physical properties of external planetary systems must necessarily 
be highly sensitive to the dynamical interactions of stellar astrospheres
with ambient interstellar matter.  This sensitivity is demonstrated by
spacecraft observations of the best-observed example, the solar heliosphere.
The space motions of nearby stars demonstrate that nearby planetary
systems will have dramatically variable interplanetary environments,
depending on the location of the star and the stellar trajectory 
relative to nearby interstellar clouds (Fig. 1).

\begin{center} {\bf References} \end{center}

\begin{itemize}
\item
"The Galactic Environment of the Sun"
Frisch, 1999,   1999, J. Geophysical Research - Blue, in press,
\item
"Dust in the Local Interstellar Wind"
P. Frisch, J. Dorschner, J. Geiss, M. Greenberg, E. Gruen, P. Hoppe, A. Jones,
W. Kraetschmer, M. Landgraf, T. Linde, G. Morfill, W. Reach, J. Slavin, J.
Svestka, A. Witt, G.Zank,   1999, Ap. J., October 20, 1999 issue
\item
"Consequences of a Change in the Galactic Environment of the Sun"  
G. Zank and P. Frisch , 1999, Ap. J., 518, 596.
(June 20, 1999)  
\item
"Galactic Environments of the Sun and Cool Stars"
Frisch, P.  1998a, In press,  Planetary Sciences - The Long View, eds. L. M.
Celnikier and J. Tran Than Van, Editions Frontieres  
\item
"The Local Bubble, Local Fluff, and Heliosphere" 
Frisch, P.  1998b, in  The Local Bubble and Beyond, Proceedings of IAU Colloquium
No. 166  (Springer-Verlag, Eds. Dieter Breitschwerdt and Michael Freyberg)
Lecture notes in Physics Series, 506, 305               
\item
"Characteristics of Nearby Interstellar Matter"  
Frisch, P.  1995, Space Science Rev., 72, 499
\item
"G Star Astropauses -- A Test for Interstellar Pressure"  
Frisch, P.  1993, Ap. J., 407, 198
\end{itemize}

\end{document}